**A method to estimate well flowing gas-oil ratio and composition using pressure and temperature measurements across a production choke, a seed composition of oil and gas, and a thermodynamic simulator**

**Seok Ki Moon, Milan Stanko. Department of Geosciences. Norwegian University of Science and Technology.**

**Corresponding author: milan.stanko@ntnu.no**

**Abstract**

In this work we propose and demonstrate a method to estimate the flowing gas-oil ratio and composition of a hydrocarbon well stream using measurements of pressure and temperature across a production choke. The method consists of using a numerical solver on a thermodynamic simulator to recombine a seed oil and gas until the simulated temperature drop across the choke is equal to the measured value. This method is meant for cases where it is not possible to measure periodically individual well composition (e.g. subsea wells producing to a common pipeline). The method uses a seed composition of oil and gas that could come from e.g. sampling and fluid testing when the well was drilled, periodic sampling of commingled well streams at processing facilities, reservoir simulation predictions, etc.

A study case and reference solution were generated using the reservoir model presented in the SPE (Society of Petroleum Engineers) comparative case Nr. 5 linked with a process simulator. Time profiles of well producing gas-oil ratio, wellstream compositions, compositions of surface conditions oil and gas, and temperature drop across the choke were generated with the models. The method proposed was then employed to estimate the flowing gas-oil ratio of the reference solution. The seed compositions employed were compositions of surface oil and gas taken from the reference solution at initial time and selected time steps (respectively representing the cases of sampling and fluid testing when the well was drilled and periodic sampling of commingled well streams in the field processing facilities).

Results show that the proposed method predicts with reasonable accuracy (maximum 12% percent error) the well gas-oil ratio and compositions during the life of the field when using compositions of surface oil and gas from initial time. When using compositions of surface oil and gas from later times, the prediction accuracy of the gas-oil ratio improves at those times but worsens for times before and after. A measurement error for the temperature drop across the choke of at least 0.01 °C is required to achieve convergence of the method. The mean percent error between the predicted and real mole fractions has an upper bound in time of 21% when using initial surface oil and gas as seed compositions.

The proposed method can be integrated in virtual metering schemes and choke models to determine surface and local flowing ratios of oil and gas. The proposed method could be also used to predict flowing water cut.

# 1. Introduction

## 1.1. Estimation of well composition

The properties of reservoir fluids are mainly dependent on their chemical (e.g. molar) composition. Accurate assessment of fluid composition is crucial for reserve estimation, field development and production facilities design, production monitoring, history matching and forecast. Lack of information about reservoir fluids may result in inappropriate design and operational choices that could bring negative consequences and financial losses to the project. Therefore, in most petroleum engineering work processes it is essential to have a good estimate of reservoir fluid properties. In the present study we focus mainly on molar composition of wellstream fluids.

The molar composition of the wellstream will change with time depending on the type of processes occurring in the reservoir (e.g. gas coning into the well, preferential mobility of oil or gas towards the well, oil vaporization, condensation, etc.).

One common method to determine the wellstream fluid characteristics is fluid sampling during well test. Reservoir fluid can be collected either at the well bottom-hole or at the surface test separator. Bottom-hole samples can be collected using wireline and high-pressure containers while surface samples can be taken from the oil and gas streams of the test separator and recombined. Fluid samples can be retrieved in exploration, at production startup, and during production. The characteristics of hydrocarbon fluids are then determined by laboratory analysis, where the list of fluid chemical components and their amounts are reported after a series of tests.

However, fluid sampling often involves significant costs. Taking bottom-hole samples may require to shut-in producing wells which causes loss of revenue. In addition, it is difficult to obtain large sample volumes due to the limited size of the sampling container and the sample might not be representative. Alternatively, surface samples are relatively easy to collect but it sampling individual wells may require installing test flowlines and a test separator. Then fluid samples of each well can be obtained by changing valve routing. However, this can be cost-prohibitive and impractical for large onshore oil and gas fields with multiple wells and commingling production network, and for subsea fields. In many cases it is not possible to measure the flowing composition during the life of the field, except for samples taken at an early stage during exploration or production drilling.

Changes in wellstream composition usually reflect on changes in the producing gas-oil ratio of the well. The gas-oil ratio in this context is the ratio between the volumes of gas and oil at standard conditions. In the past, authors have proposed methods to estimate fluid compositions using gas-oil ratio measurements and recombination of "seed" oil and gas in different proportions until the measured gas-oil ratio is achieved. The recombination is typically performed experimentally in the laboratory or with the use of numerical simulation.

The producing oil-gas ratio also depends on the surface process, e.g. number of stages and pressures and temperatures of each stage. Moreover, the gas-oil ratio also depends on the location in the separation facilities where the oil and gas rates are measured.

The choice of the seed oil and gas composition has an important effect in the results of the recombination process. Thomas et al. (2007) discussed three methods to estimate reservoir composition using seed compositions of oil and gas. They highlight that a challenge when using separator oil and gas as seed composition is that the composition obtained might have a saturation pressure that is higher than that of reservoir fluids. This is because separator gas is usually rich in light hydrocarbon components due to the high mobility of gas towards the wellbore and potential gas coning from the gas cap. They discuss workarounds such as “depleting” initial reservoir composition to current reservoir pressure and use the resulting oil and gas as seed compositions and using the gas in the gas cap as an additional source of gas seed composition.  A similar method is discussed in Sthener et al (2010), in Fig. 4 where they “deplete” separately the initial oil and gas layers to current reservoir pressure. They then recombine the resulting oil and gas in the oil layer with the resulting gas from the gas layer. The proportion between oil and gas from the initial oil layer was bounded considering the physically possible variation of oil and gas mobilities in the reservoir.

Commercial simulators usually have facilities to estimate a wellstream composition with the producing gas-oil ratio.  Some examples are the simulator PVTSim (as reported by Hoda and Whitson 2013) and the well simulator Prosper (Petroleum Experts 2019).

Hoda and Whitson (2013) highlight that separator conditions might vary widely in a field and therefore there are no consistent reference conditions to report surface oil and gas rates and oil-gas ratios. Therefore, gas and oil surface rates measured in one well cannot be directly compounded with surface rates measured in another well.  They proposed a method, requiring no iteration, to convert volumetric well test rates to compositional wellstream using recombination of separator (or surface) oil and gas. This compositional wellstream can then be passed by a reference surface process (using a numerical simulator) to compute surface rates of oil and gas and gas-oil ratios.

### 1.2. Flow through production choke

The most common purpose of a production choke valve is to control the well production rates of oil, gas and water by reducing the pressure of the stream. It consists of passing the fluid flow through a restriction, sometimes referred to as “throttling” process. When fluid encounters the contraction point, the fluid flowing pressure drops as the fluid velocity increases. After the contraction point the channel expands again to pipe size, fluid velocities are reduced, but the pressure does not return to its upstream levels due to friction and localized energy losses.

The flow of multiphase flow through restrictions has been extensively investigated by several researchers both theoretically and experimentally. The research aim is often the prediction of mass flow passing through the production choke valve using the pressure drop. Gilbert (1954), Ros (1960), Omana et al. (1969), Henry and Fauske (1971), Ashford (1974) presented experimental investigation on multiphase flow through the restriction and presented empirical correlations between the pressure, flow rate and the size of the restriction. Subsequently, Sachdeva et al. (1986), Perkins (1993), Schüller et al. (2003, 2006), Al-Safran and Kelkar (2009) developed mechanistic models based on mass and momentum balance equations under the assumption of the polytropic flow. More recent research is by Al Ajmi et al. (2015), Guo et al. (2002), Hong et al. (2018).

Authors have approximated the thermodynamic process across a choke (or across parts of the choke, e.g. between inlet and contraction) as adiabatic (Ros, 1960), isothermal, isentropic, isenthalpic or polytropic (Al-Safran & Kelkar, 2009). Sometimes different thermodynamic processes are used for the gas and liquid and mass transfer is often considered negligible.

The isenthalpic assumption is often justified by arguing that the choke is relatively small, ~~the fluid velocity is high~~ and there is no significant heat transfer with the environment. Additionally, there is no work exchanged during the expansion with the environment and the change in potential and kinetic energy between inlet and outlet is negligible.

### 1.3. Research gap and goal of study

The challenge and research gap addressed in this study is estimating well composition (and flowing gas oil ratio) for cases in which periodic well sampling is impractical, costly or impossible (e.g. subsea wells producing to a common pipeline). The solution proposed builds up on existing methods that rely on recombination of seed oil and gas, and using temperature and pressure data upstream and downstream a production choke.

In this work we propose a method to predict the gas-oil ratio of a wellstream combining two elements presented earlier:

- Assume that the expansion across the choke is isenthalpic and that upstream and downstream pressure and temperatures are known
- The wellstream fluid can be approximated by combining a seed composition of oil and gas

The method uses a thermodynamic simulator. We first obtain seed compositions of oil and gas by passing an initial composition (from e.g. sampling and fluid testing when the well was drilled) through the surface process. We then use a numerical solver on the simulator to find the proportion of seed oil and gas required to obtain the measured downstream temperature. This assuming an isenthalpic expansion from the measured upstream pressure and temperature to the measured downstream pressure. The wellstream composition obtained is then passed by the surface process to calculate its surface gas-oil ratio.

## 2. Study case

Figure 1 shows a schematic of the production system used in this study. The wellstream flows through a production choke valve with an ~~input~~ pressure drop, inlet pressure and inlet temperature. Thereafter, it flows through the pipeline until it reaches the surface process, which is a 2-stage separation process. At the outlet of the surface process, the output oil and gas are taken to standard conditions. The pressure and temperature on both upstream and downstream side of the production choke are known.

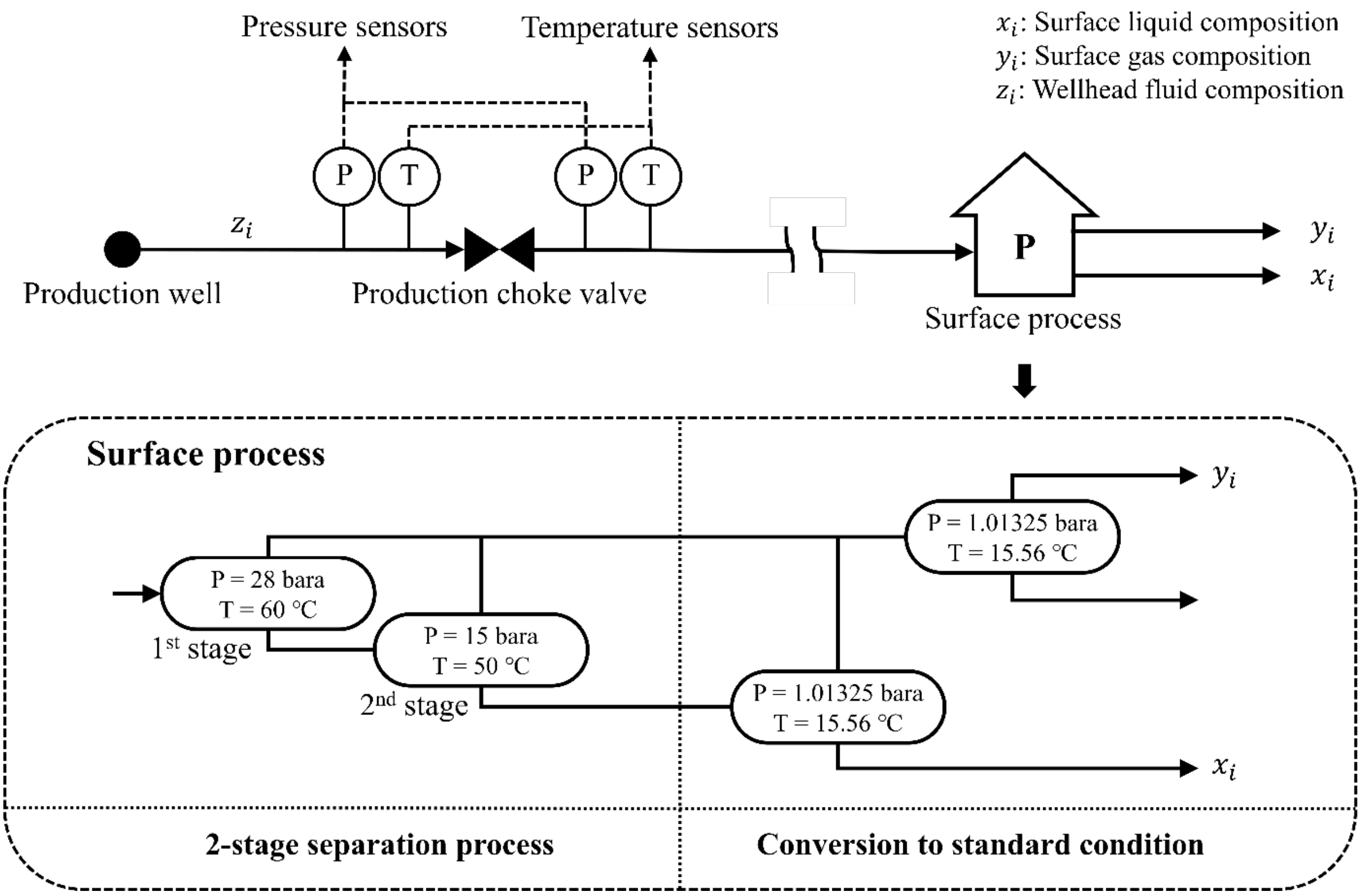

**Fig. 1. Schematic of production system.**

### 2.1. Calculation of the reference solution (well composition, choke discharge temperatures and flowing gas oil-ratio for all time steps)

The reference solution is generated using two simulators, a commercial compositional reservoir simulator (SENSOR, April 1, 2011, Coats Engineering, Inc.) and a commercial process simulator (Aspen HYSYS, V10, Aspentech). The reservoir model used is the SPE comparative case Nr. 5 (presented in Killough & Kossack, 1987). The model was modified to remove injection and to use the surface process shown in Fig. 1. The model has a single well completed in a single cell and is run for a simulation time of 7.6 years, with a well oil target rate of 1908 $Sm^3/d$. More details about the model are provided in Appendix A. More details can also be found in the work by Killough & Kossack, 1987.

The resulting profiles of wellstream composition ($z_i$) and pressure upstream the choke in time were then input to the process simulator. The process simulator employs the Peng-Robinson equation of state and liquid density is computed by Costald method.  The process simulator models the isenthalpic choke expansion, the separation process, and the conversion to standard conditions. The choke expansion uses as input a constant pressures drop of 30 bara, and a constant inlet temperature of 66 °C. The piping between wellhead and surface process is not modeled.

The process model was then run for all time steps of the solution of the reservoir simulator. Values computed are the time profiles of downstream temperature of the choke, the producing gas oil ratio (shown in Fig. 2) and the composition of standard conditions gas ($y_i$) and oil ($x_i$). The temperature drop across the choke versus the gas-oil ratio is shown in Fig. 3. It can be seen that the magnitude of the temperature difference is modest, ranging from -1.5 to 4 °C. For most GOR values, there is a unique relationship between temperature difference and GOR, except for the early ~~and late~~ stages in the life of the field where GOR is below ~~200~~ 150 Sm$^3$/Sm$^3$.

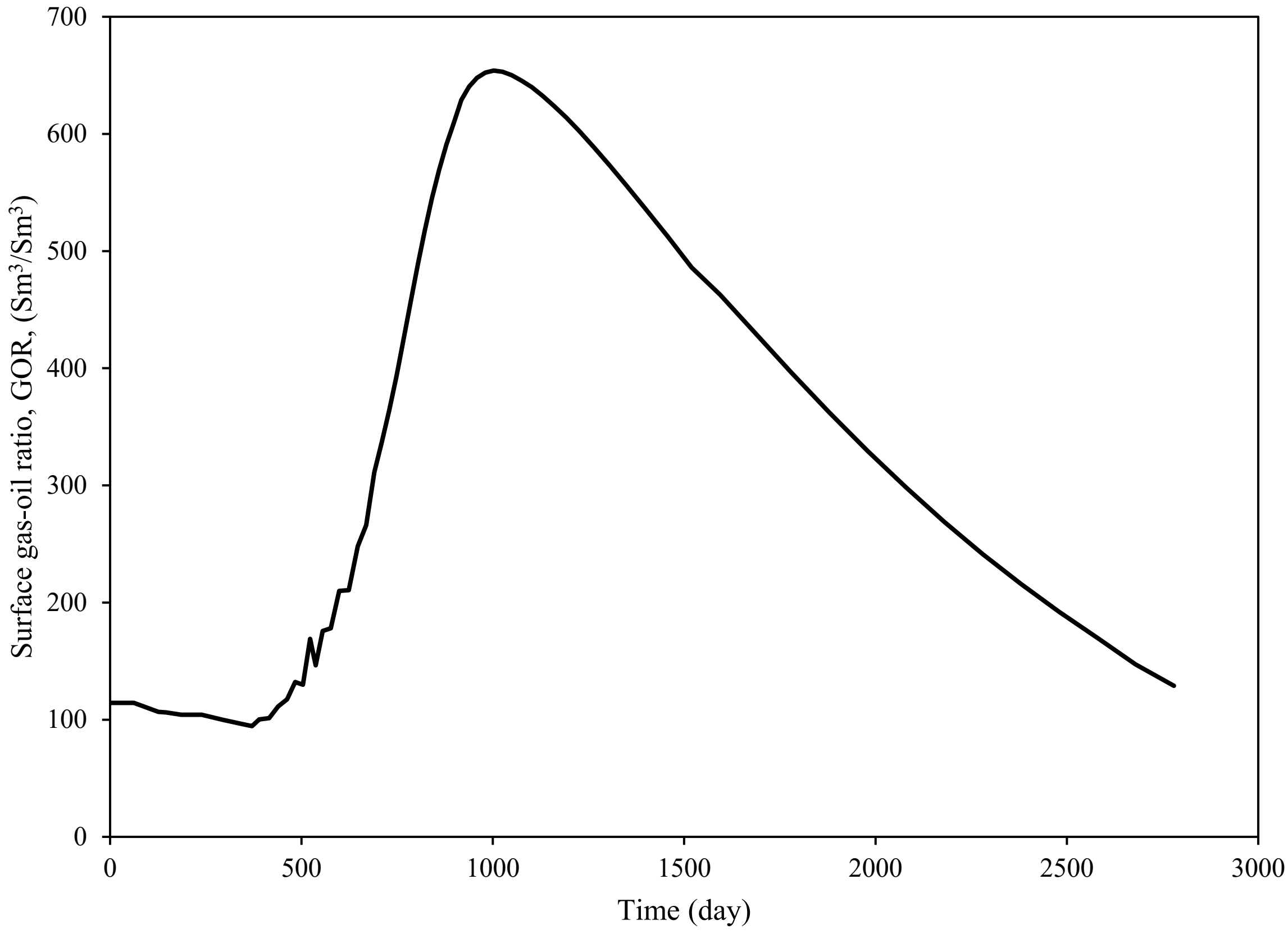

**Fig. 2. Reference case surface GOR**

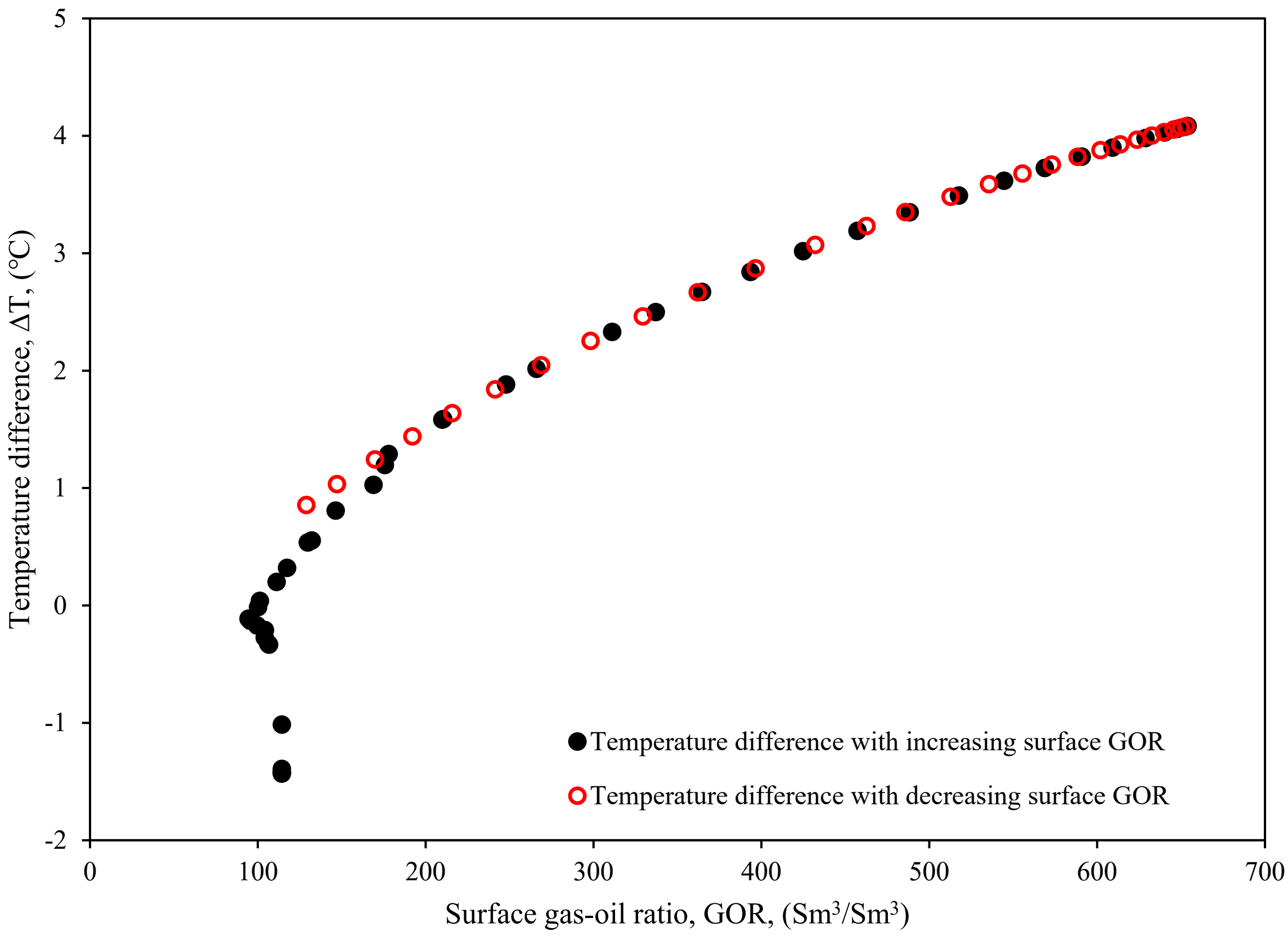

**Fig. 3. Temperature difference across the production choke valve by reference case surface GOR.**

The results described in this section are the "true solution" that the proposed method attempts to approximate. In a real oil and gas field, the upstream and downstream choke pressures and temperatures are usually available from measurements. The composition of surface oil and gas of commingled well streams ~~or, alternatively, reservoir composition,~~ is usually available from periodic sampling at the processing facilities. In this case it is assumed it is not possible to measure periodically individual well composition (e.g. subsea wells producing to a common pipeline). A well composition could be available e.g. from sampling and fluid testing when the well was drilled.

## 3. Concept and workflow

Figure 2 presents the proposed simulation-based workflow to predict wellstream composition using "measured" values of temperature downstream the production choke and seed compositions of oil and gas. In the workflow, a surface gas and surface oil of compositions ($y_i$) and ($x_i$) respectively are recombined in a specific proportion (in Fig. 4, using the variable $f_g$) to compute the wellstream composition $z_i$ . An isenthalpic flash is then performed from measured inlet pressure and temperature to a measured outlet pressure. This process represents wellstream fluids passing through the production choke. Fluids can be a single phase or multiphase fluid depending on the pressure and temperature conditions.

When the fluid stream passes through the production choke, a downstream temperature is computed. The calculated temperature is then compared against the measured value. If equal, this means the wellstream composition $z_i$ is a good approximation of the true composition. If not, the recombination ratio should be adjusted until both calculated and measured temperatures are equal. This can be conducted iteratively until a numerical solver. Once convergence is attained, the resulting composition $z_i$ is passed through the surface process to obtain the producing gas-oil ratio.

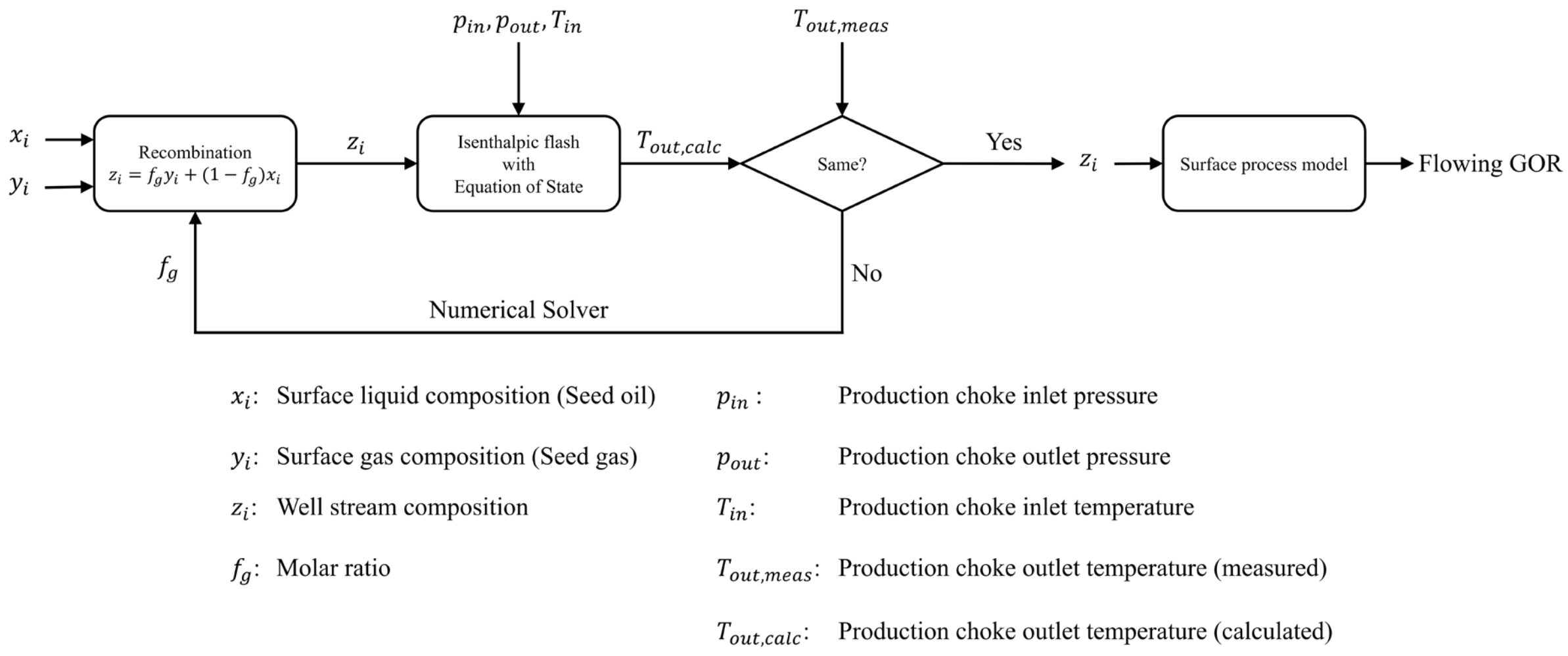

**Fig. 4. Workflow of the proposed concept.**

The following tolerance values of the numerical solver were tested: 0.1, 0.01, 0.001 and 0.0001 °C. The goal of using several tolerances was to study the effect the accuracy of the temperature sensor has on the accuracy of the method.

The seed compositions of oil and gas were obtained using compositions of surface oil and gas from the reference solution for some specific days, namely day 2, 370, 598, 690, 748, 803 and 1002 (from production startup to the time where producing GOR is maximum). The goal of using different seed compositions was to evaluate how does the accuracy of the approximation varies if more recent samples of surface oil and gas are available. Compositions of surface oil and gas of day 2 represents there is only one sample available, ~~taken at production startup~~ e.g. taken when the well was drilled. Compositions of surface oil and gas at later days represent periodic sampling of commingled well streams at the processing facilities.

## 4. Results

Figure 5 shows the time profile of the reference solution GOR and the GOR estimated with the method using as seed compositions surface oil and gas of day 2. The agreement between them is fair, accurate at the beginning and the end of the production while higher (12% percent error) when the GOR is maximum. This could be because the day 2 wellstream fluid is a single-phase liquid rich in heavy components. Thus, the surface gas used as seed composition will also have high amounts of heavy components. In contrast, the surface gas at the maximum GOR has fewer heavy components. The light green curve in Fig. B.1 (Appendix B) shows the percent error of the estimated surface GOR.

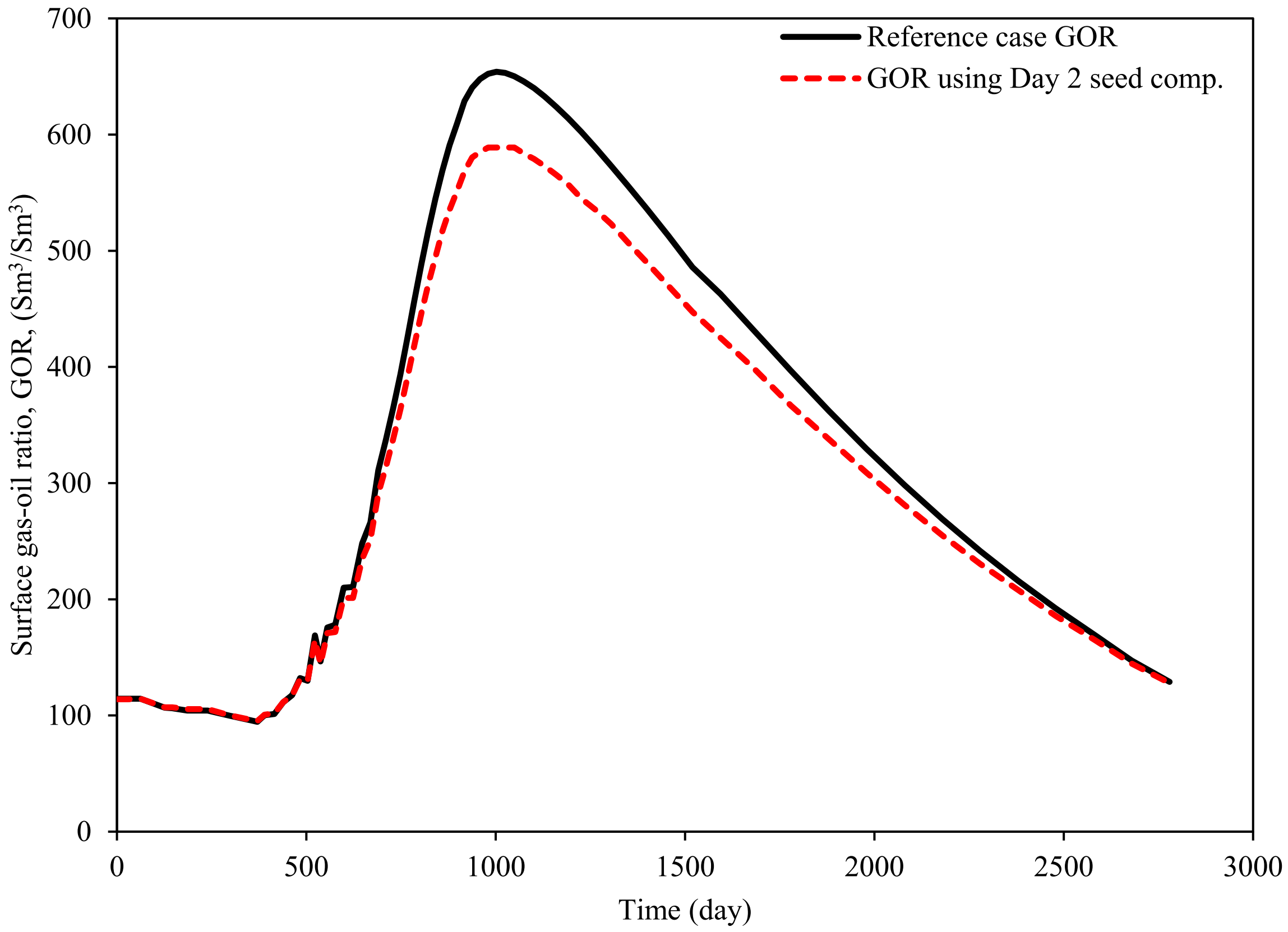

**Fig. 5. Original surface GOR and estimated surface GOR with Day 2 seed composition.**

### 1.1. Effect of solver tolerance

Figure 6 shows the estimated producing GOR using solver tolerances of 0.1, 0.01, 0.001 and 0.0001 °C. For tolerances equal to and below 0.01 °C all estimated profiles of GOR overlap, meaning that the method is converged.

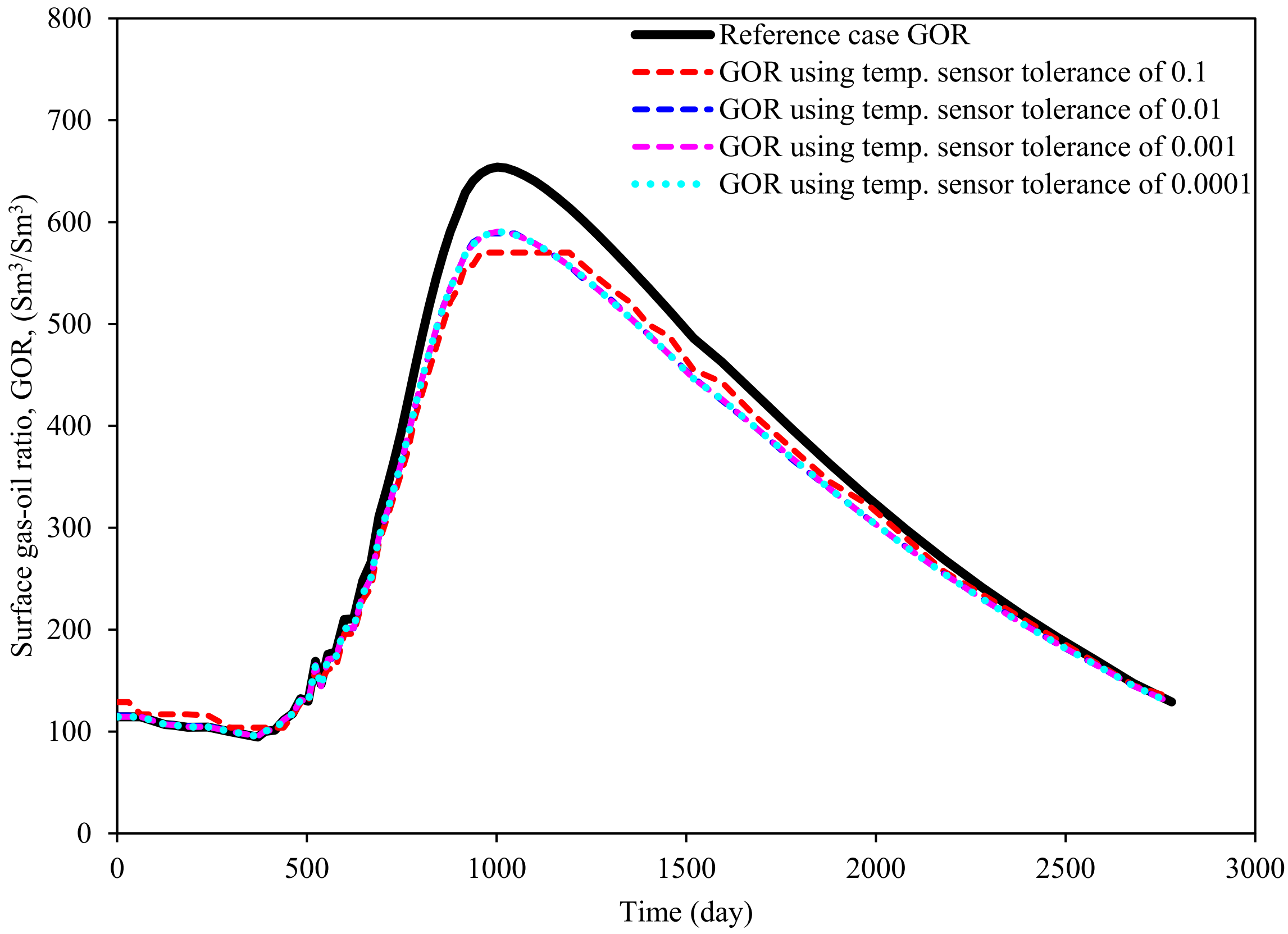

**Fig. 6. Estimated surface GOR with four different temperature sensor tolerances.**

Figure 7 shows the percent error of estimated surface GOR with different temperature sensor tolerances. For tolerance values equal or below 0.01 °C, the curves of GOR percent error overlap, however, for 0.01 °C some minor fluctuations are exhibited.

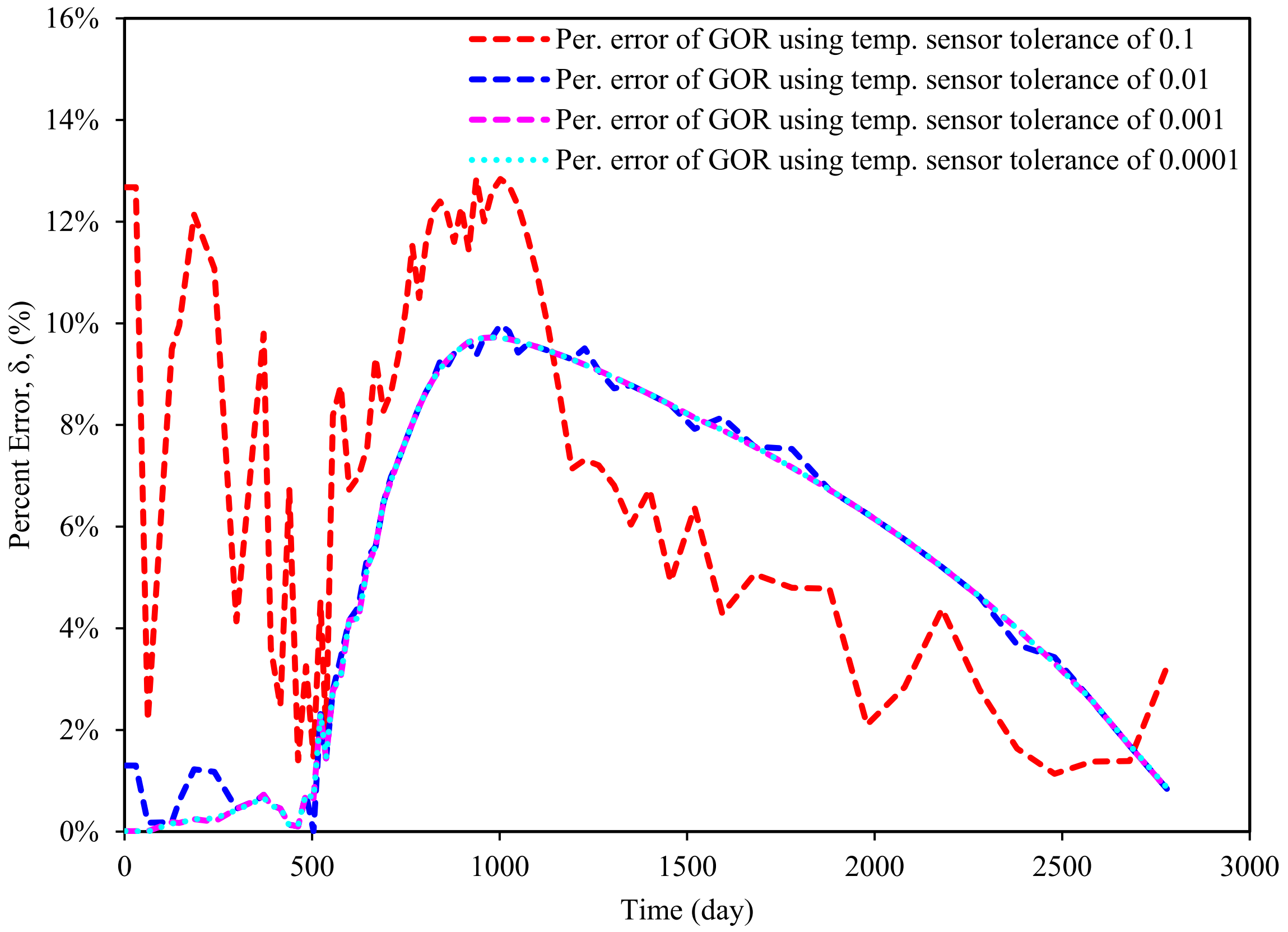

**Fig. 7. Percent error of estimated surface GOR with four different temperature sensor tolerances.**

Figure 8 shows the mean percent error between the reference solution and the approximated mole fraction for all six components, for different values of solver tolerance. The results are similar to the ones shown in Fig 7, i.e. convergence is achieved for tolerance values equal to or lower than 0.01 °C. For subsequent study cases presented in this work, the solver tolerance used is of 0.01 °C.

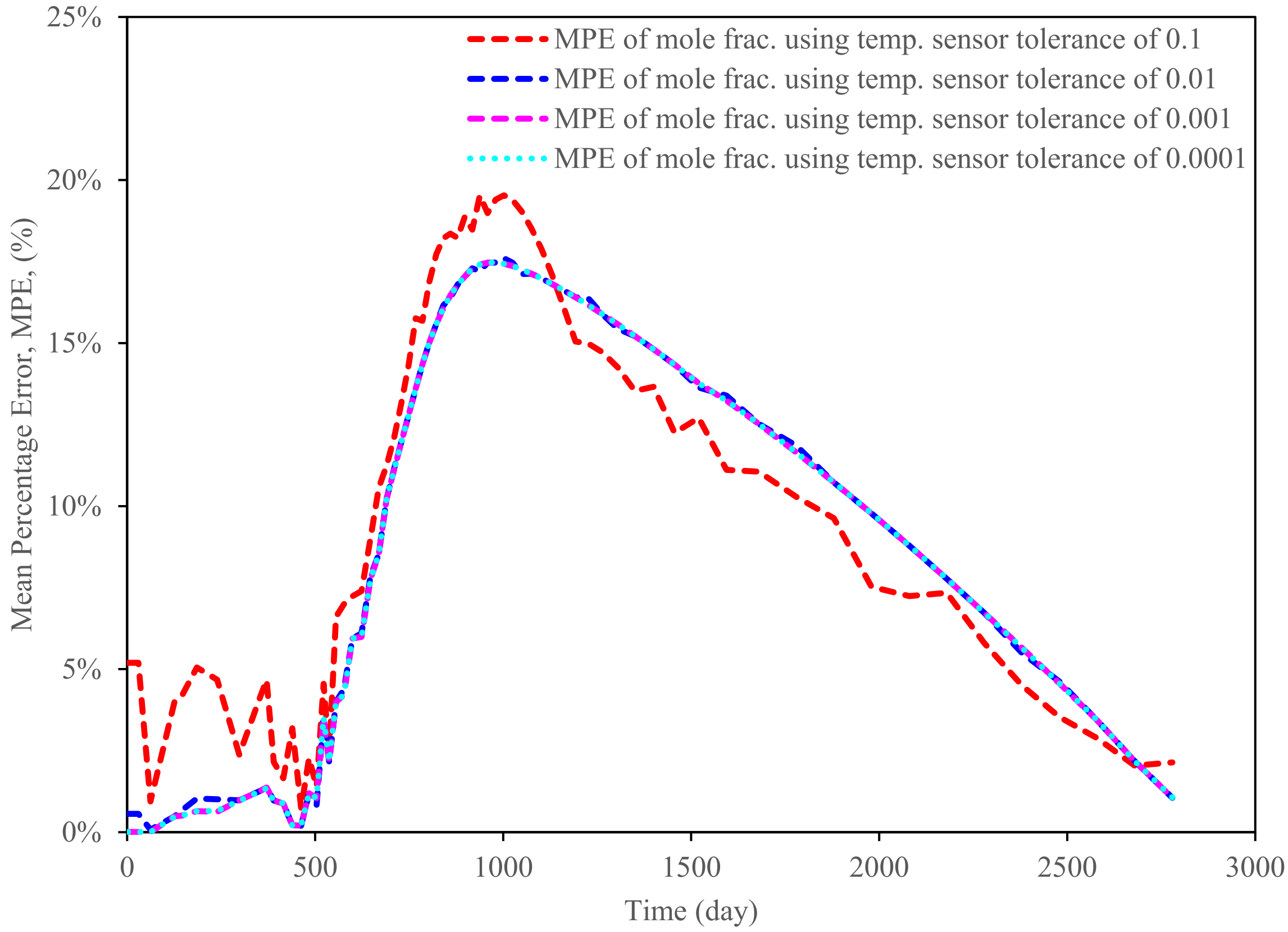

**Fig. 8. Mean percent error of mole fraction of each six components with four different temperature sensor tolerances.**

### 4.1. Effect of seed composition selection

Figure 9 shows the estimated surface GOR over the entire production period using as oil and gas seed compositions surface oil and gas from the reference solution at seven different time points. The accuracy of the estimated surface GOR depends strongly on the seed compositions. It seems that the accuracy is improved significantly if the seed composition employed is the surface oil and gas from the time where the GOR is maximum.

Figure B.1. (in Appendix B) shows the percent error of estimated surface GOR for the cases shown in Fig. 9. When using seed compositions taken from later days (i.e. closer to the time at which the maximum GOR is registered), the accuracy of the prediction is improved at those days, but it worsens at early and late days.

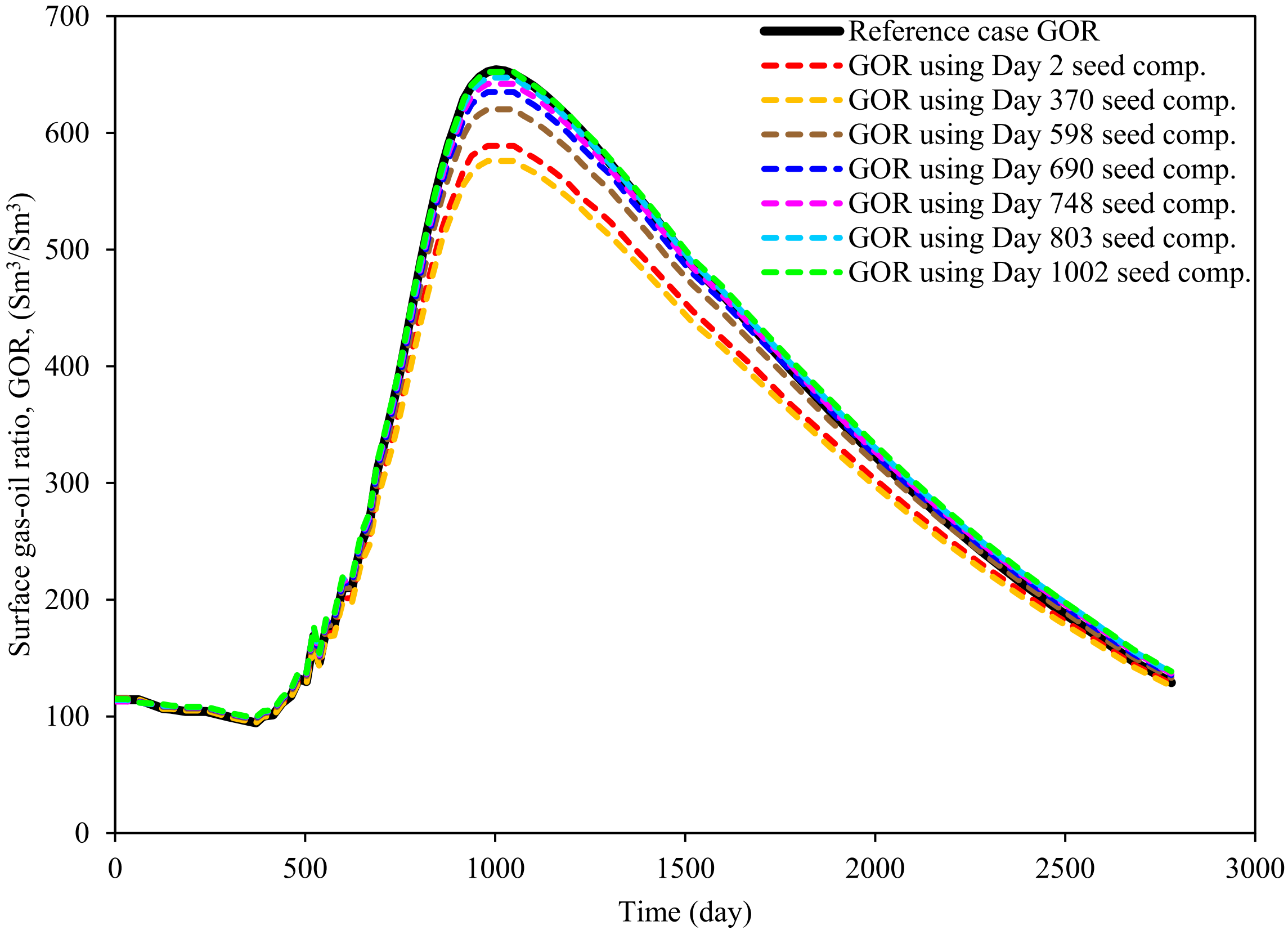

**Fig. 9. Original surface GOR and estimated surface GOR with different seed compositions of seven time points.**

These results indicate that the method proposed could be improved by updating the seed compositions of oil and gas in time whenever new information is available (e.g. sampling is performed).

Figure 10 shows the mean percent error of mole fraction of all components between the estimated and reference solution composition. The behavior is similar to the percent error on the GOR, low at initial and late times, and high when close to times where the GOR is maximum. The same effect of seed composition selection is observed: when the seed composition is taken from surface oil and gas of later times, the percent error improves at those times, but it increases at early and late times.

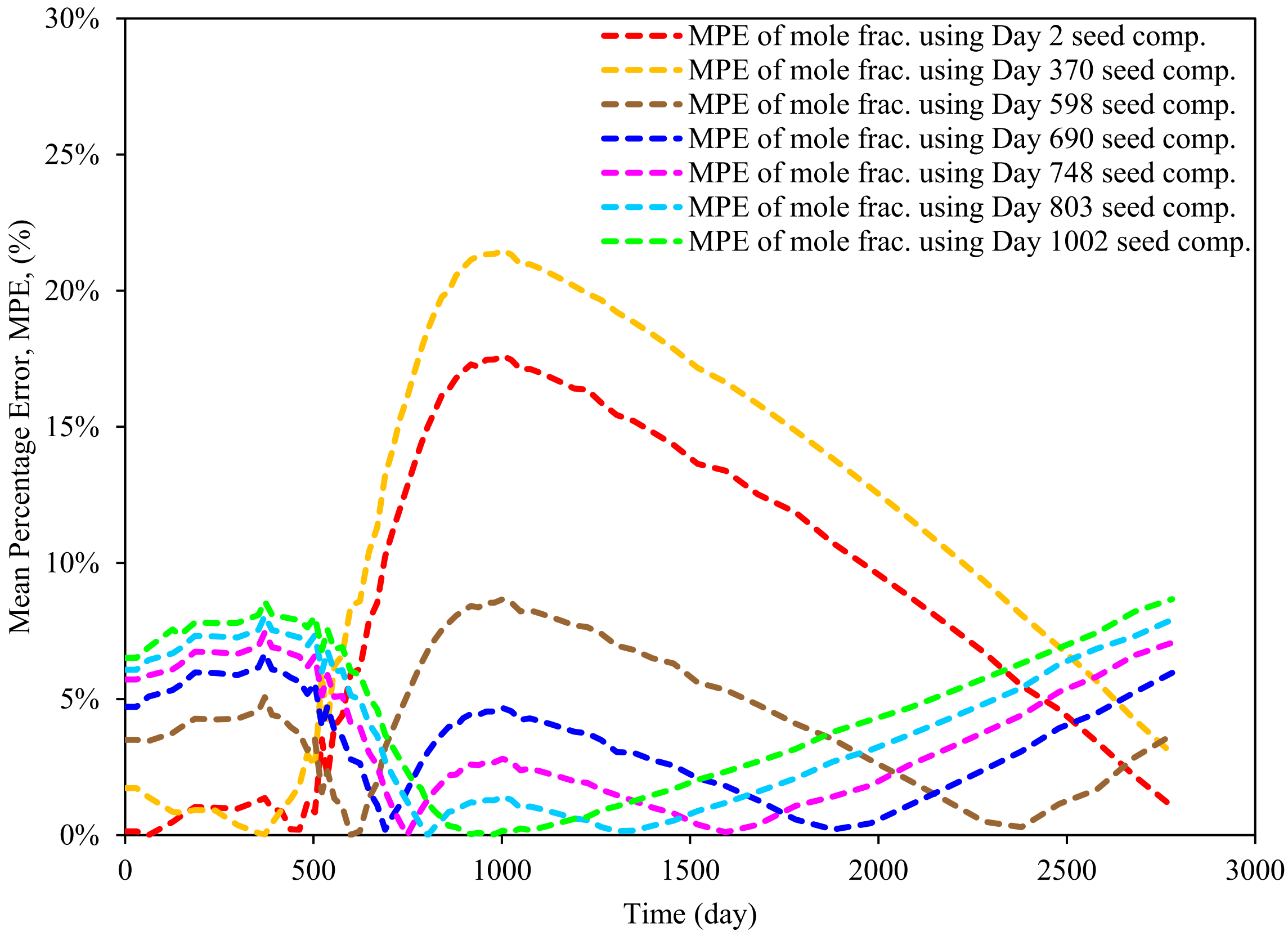

**Fig. 10. Mean percent error of mole fraction of each six components with different seed compositions of seven time points.**

## 5. Conclusion

- This paper presents a method to estimate wellhead fluid compositional stream by using measured pressure and temperature difference across the production choke valve, a seed composition of oil and gas and a thermodynamic simulator. The method employs a numerical solver on the thermodynamic simulator to find the proportion of seed oil and gas that provides a simulated temperature at the choke outlet equal to the measured. The expansion in the choke is assumed to be isenthalpic.
- Seed compositions used in this work were surface oil and gas taken from the reference solutions at selected time steps, representing the case of sampling and fluid testing when the well was drilled and periodic sampling at the processing facilities of commingled well. The accuracy of the method to approximate the wellstream GOR and composition are high at those times from where the surface oil and gas are taken from and worsens gradually for times before and after. However, the accuracy of the method is fair if seed composition is taken from surface oil and gas at initial time, the highest percent error registered is of 12 % for the GOR and 25% for the mole fraction.

- The required tolerance of the numerical solver to achieve convergence must be equal or lower than 0.01 °C. This means that the accuracy of the temperature measurement downstream the choke must also be equal or lower than 0.01 °C to ensure method convergence.
- The proposed method could support virtual metering schemes in production systems where fluid sampling is unfeasible or expensive.
- The proposed method can be used to determine beforehand the optimal fluid sampling interval to update seed compositions of oil and gas.
- The proposed method can be applied to determine flowing water cut or condensate water ratio instead of gas oil ratio.

## 6. Nomenclature

$x_i$: Mole fraction of component “i" in surface liquid composition (Seed oil)

$y_i$: Mole fraction of component “i” in surface gas composition (Seed gas)

$z_i$: Mole fraction of component “I" in wellhead fluid composition

$f_g$: Molar ratio between molar rate of seed gas and total molar rate (seed oil and seed gas)

$p_{in}$: Production choke inlet pressure (bara)

$p_{out}$: Production choke outlet pressure (bara)

$T_{in}$: Production inlet temperature (°C)

$T_{out,\,meas}$: Production choke outlet temperature (°C, measured)

$T_{out,\,calc}$: Production choke outlet temperature (°C, calculated)

δ: Percent error

GOR: Gas-oil ratio

MPE: Mean percentage error

## 7. Acknowledgments

We would like to thank Mohamad Mazjoub Dahouk from Whitson AS for performing the reservoir simulations and Coats Engineering for providing an academic license of their reservoir simulator. Also, this research is a part of BRU21 – NTNU Research and Innovation Program on Digital and Automation Solutions for the Oil and Gas Industry (www.ntnu.edu/bru21) and supported by AkerBP.

## APPENDIX A

### A.1 Water properties

| Item | Value | Unit |
|---|---|---|
| Initial water formation volume factor | 1.0 | rb/stb |
| Water compressibility | 3.3E-6 | 1/psi |
| Stock tank water density | 63.22 | lbs/ft$^3$ |
| Water viscosity | 0.7 | cp |
| Rock pore volume compressibility | 5.0E-6 | 1/psi |
| Reference pressure for water volume factor | 3984.3 | psia |

### A.2 Equation of state PVT data

| Item | Value | Unit |
|---|---|---|
| Equation of state | Peng-Robinson | - |
| Reservoir temperature | 160 | °F |

A.3.1. PVT properties by components

| Component | Critical pressure | Critical temperature | Molecular weight | Parachor | Acentric factor | Critical z-factor |
|---|---|---|---|---|---|---|
| | psia | °R | - | - | - | - |
| C1 | 667.8 | 343 | 16.04 | 71 | 0.011 | 0.286 |
| C3 | 616.121 | 665.7 | 44.1 | 151 | 0.1524 | 0.277 |
| C6 | 438.74 | 913.68 | 86.18 | 271 | 0.297 | 0.264 |
| C10 | 306.0. | 1111.86 | 142.29 | 431 | 0.491 | 0.256 |
| C15 | 214.656 | 1274.4 | 212.419 | 631 | 0.685 | 0.243 |
| C20 | 168.244 | 1380.6 | 282.547 | 831 | 0.9065 | 0.213 |

A.3.2. Binary Interaction coefficients

| | C1 | C3 | C6 | C10 | C15 | C20 |
|---|---|---|---|---|---|---|
| C1 | 0 | 0.119 | 0.04 | 0.0489 | 0.0489 | 0 |
| C3 | | 0 | 0.0007 | 0 | 0 | 0 |
| C6 | | | 0 | 0 | 0 | 0 |
| C10 | | | | 0 | 0 | 0 |
| C15 | | | | | 0 | 0 |
| C20 | | | | | | 0 |

### A.3 Relative permeability data

A.3.1. Two-phase water-oil saturation table

| $S_w$ | $K_{rw}$ | $K_{row}$ | $P_{cwo}$ |
|---|---|---|---|
| 0.2 | 0 | 1 | 45 |
| 0.2899 | 0.0022 | 0.6769 | 19.03 |
| 0.3778 | 0.018 | 0.4153 | 10.07 |
| 0.4667 | 0.0607 | 0.2178 | 4.9 |
| 0.5556 | 0.1438 | 0.0835 | 1.8 |
| 0.6444 | 0.2809 | 0.0123 | 0.5 |
| 0.7 | 0.4089 | 0 | 0.05 |
| 0.7333 | 0.4855 | 0 | 0.01 |
| 0.8222 | 0.7709 | 0 | 0 |
| 0.9111 | 1 | 0 | 0 |
| 1 | 1 | 0 | 0 |

A.3.2. Two-phase gas-oil saturation table

| $S_{liq}$ | $k_{rg}$ | $k_{rog}$ | $P_{cgo}$ |
|---|---|---|---|
| 0.2 | 1 | 0 | 30 |
| 0.2889 | 0.56 | 0 | 8 |
| 0.35 | 0.39 | 0 | 4 |
| 0.3778 | 0.35 | 0.011 | 3 |
| 0.4667 | 0.2 | 0.037 | 0.8 |
| 0.5556 | 0.1 | 0.0878 | 0.03 |
| 0.6444 | 0.05 | 0.1715 | 0.001 |
| 0.7333 | 0.03 | 0.2963 | 0.001 |
| 0.8222 | 0.01 | 0.4705 | 0 |
| 0.9111 | 0.001 | 0.7023 | 0 |
| 0.95 | 0 | 0.88 | 0 |
| 1 | 0 | 1 | 0 |

The SPE comparative reservoir case Nr. 5 employs six pseudo-components and their characteristics such as critical pressure, temperature and Z factor, molecular weight and acentric factor are provided as input. However, for the calculation of enthalpy, other input parameters for each component are required that are not available in the original publication by Killough & Kossack, 1987. Therefore, it was decided to use in the process simulator normal alcanes from the database with the same carbon number rather than the pseudo components. This does not cause any consistency issues because the reservoir simulator is used to compute the wellstream composition in time, while the choke expansion, surface process and conversion to surface conditions is simulated in the process simulator.

**APPENDIX B**

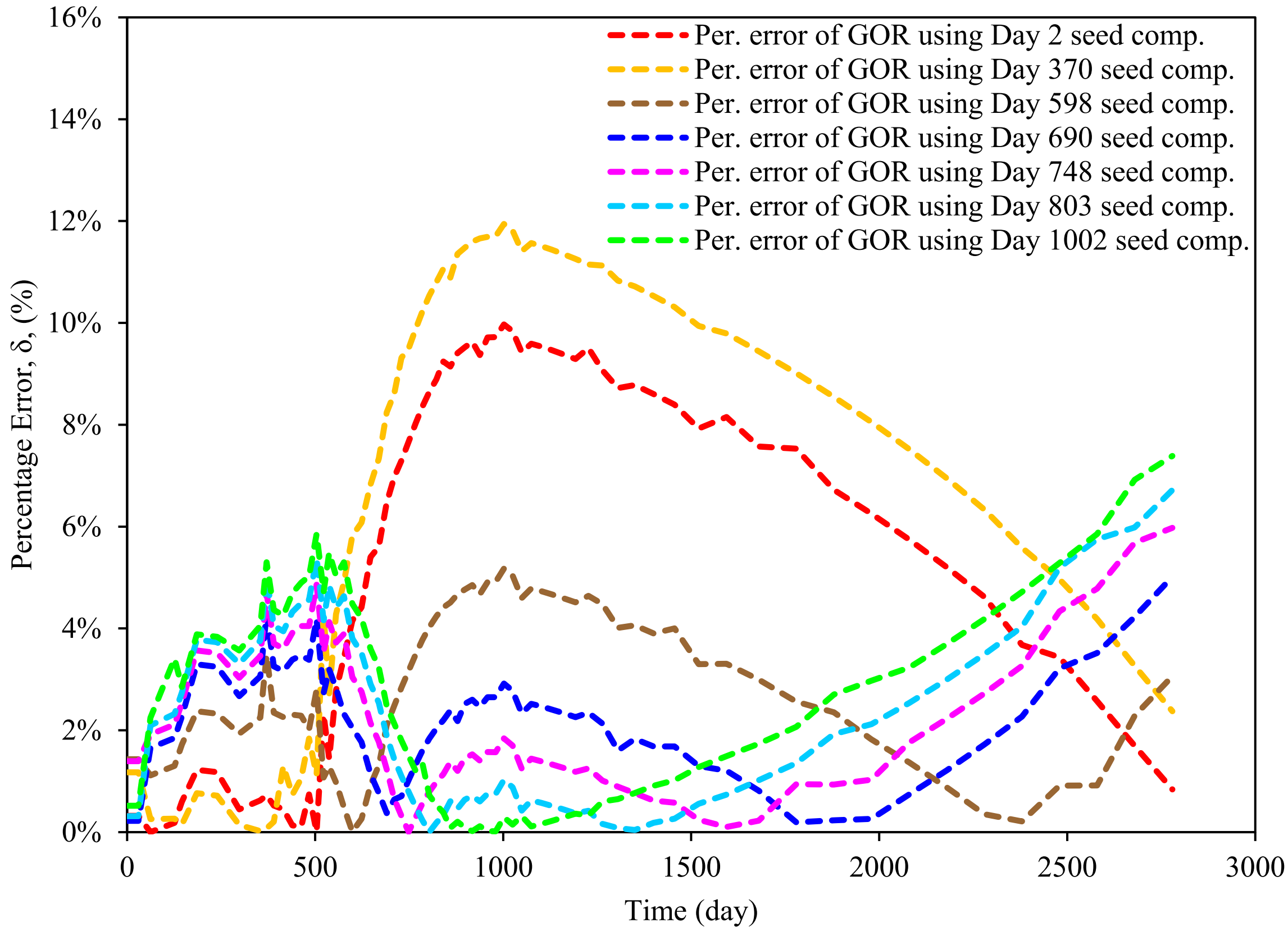

**Fig. B.1. Percent error of estimated surface GOR with different seed compositions of oil and gas taken at seven time steps from the reference solution.**

Trondheim, October 2nd, 2024.

**Introduction by Milan Stanko.**

This manuscript was originally submitted to the Elsevier Journal of Petroleum Science and Engineering[1] by Mr. Moon on October 21st, 2020. The manuscript number assigned was PETROL22932.

Rejection notification and reviewer comments were received on September 24th, 2020. The comments from the two reviewers are pasted next, together with answers provided by Milan Stanko (in blue color). To address some of these comments, some modifications have been performed on the original manuscript. Additions are marked in blue color and subtractions are marked in red color.

The new version of the manuscript will hopefully be published in a different journal, or conference sometime in the near future, pending action by co-author Seok Ki Moon. In the meantime, I welcome people to provide comments either to milan.stanko@ntnu.no, or directly on https://pubpeer.com/

**COMMENTS FROM EDITORS AND REVIEWERS**

**Comment by Milan:** I thank the reviewers for their time invested in providing comments to this work.

**Reviewer #1:**

This manuscript is more like a technical report and therefore I can't suggest its publication.

**Answer by Milan:** I respectfully disagree with the reviewer. This is not a strong argument against publication. Many technical reports have high scientific value. More clarifications are required about this comment.

The manuscript claims presenting a new methodology for estimating producing GOR from downstream and upstream P and T across choke and gas/oil compositions. The main issue is that in real field, obtaining fluid composition is even more expensive and difficult than producing GOR. If we can measure compositions at test separators, then at the same time we can directly measure GOR. In this regard, I do not see this methodology useful, and is redundant.

**Answer by Milan:** I respectfully disagree with the reviewer. There are many cases in which it is not possible to periodically measure current well composition, for example subsea (or onshore) clusters of wells that produce to a subsea (or onshore) pipeline. In these cases, sampling typically performed on topside facilities is of commingled well streams. Usually, only a well sample and PVT report is available from when the well was drilled. This method is useful for those cases, which are common in e.g. Norway.

[1] ISSN: 0920-4105, later continued as Geoenergy Science and Engineering, ISSN: 2949-8929

Another issue with this manuscript is novelty. Even if the presented methodology is useful, it is not novel from scientific point of view.

**Answer by Milan:** After performing literature review, we didn't find any works similar to what is proposed in the manuscript. There are many works that deal with recombination to obtain well composition (and many have been properly referenced in the manuscript), but none using choke temperature variation. But it could be we missed them. We appreciate the reviewer to provide an example or examples.

It uses commercial softwares and back calculation processes. Instead of starting from A and reaching B, this manuscript suggest starting from B and reach A. !

**Answer by Milan:** I respectfully disagree with the reviewer. The fact that we are using commercial software and back-calculation is not a serious argument against publication. Many scientific publications use commercial software. We needed a thermodynamic engine to do our study, and we used a commercial software available at our university. I don't see the point of programming our own thermodynamic engine and spend space in the manuscript explaining and validating it, since it is not the main point of the work.

Other issues:

Add page numbers

Add line numbers

**Answer by Milan:** Thanks for the comment. I apologize this was missing in the submission. Page and line numbers will be added when preparing for future submission.

Section 1.2. there are many newer correlations for flow through chokes. Authors have missed many of them.

**Answer by Milan:** It is true that there are newer choke correlations, but our work is not using choke correlations, it is referring to the fact that choke correlations use assumptions about the thermodynamic process the fluid experiences when expanding through a choke. Therefore, discussing all possible and new choke models is irrelevant to the work presented.

Literature review is almost none.

**Answer by Milan:** I respectfully disagree with the reviewer. Considering the novelty of the idea, I consider sufficient the literature presented and reviewed (17 references). It is poor practice to include many irrelevant references.

**Reviewer #2:**

A method to estimate flowing gas-oil ratio using the pressure and temperature measurement of a choke is proposed in this study. However, the manuscript is not well-written. It is hard to follow the authors' method. I cannot find the research gap and innovative point of this study. The proposed method seems very simple for me. Authors simply used some commercial software. Besides, I don't think this topic is of interest to both the industry and academic.

**Answer by Milan:** Thank you for the comment. I am sorry to hear you find the manuscript is not well-written. To prepare for future submissions we will do a language review/proofing of the text.

However, the reviewer contradicts himself/herself by emitting the following statements "It is hard to follow the author's method" and "The proposed method seems very simple for me". Which one is it?

The fact that we are using commercial software and back-calculation is not a serious argument against publication. Many scientific publications use commercial software. We needed a thermodynamic engine to do our study, and we used a commercial software available at our university. I don't see the point of programming our own thermodynamic engine and spend space in the manuscript explaining and validating it, since it is not the main point of the work.

The work could be very relevant for the Norwegian oil and gas industry, e.g. production allocation cases where the flowing gas-oil ratio and water cut of the stream is not known. This is needed for e.g. multiphase metering, virtual flow metering schemes.

More comments are as follow:

1.Authors didn't explain the proposed method in the abstract very well. They should briefly introduce the principle or theory behind their method. The descripting in the abstract is confusing.

**Answer by Milan:** Thank you for the comment. I am sorry to hear you find the abstract is confusing. We will improve it when preparing for future submissions.

2.In section 1, what is the research gap according to your literature review?

**Answer by Milan:** Thanks for the comment. I am sorry this was missing in the original version of the manuscript.

3.Besides, I am wondering what's the point of this study. It should be very easy to determine the oil-gas composition using a field sample. I would believe this is the most commonly used method in field application.

**Answer by Milan:** There are many cases in which it is not possible to periodically measure current well composition, for example subsea (or onshore) clusters of wells that produce to a subsea (or onshore) pipeline. In these cases, sampling typically performed on topside facilities is of commingled well streams. Usually, only a well sample and PVT report is available from when the well was drilled. This method is useful for those cases, which are common in e.g. Norway.

4.In section 3, the proposed method is very simple, I cannot see any innovative point of this study.

**Answer by Milan:** I respectfully disagree with the reviewer. The fact that the method is simple, is not a serious argument against publication or against its merits. In many situations, the simpler, the better. Despite its simplicity, we haven't seen this method used/published before, which I believe speaks to its novelty.

5.How do you make the initial assumption of the gas-oil composition? If you don't have a sample, there might be hundreds of different types of fluids. If you have a sample, you could easily test the composition, then you can calculate the downhole fluid composition and characteristics according to pressure and temperature.

**Answer by Milan:** In subsea wells commingled into a single pipeline, an option to get the composition of seed oil and gas is use a sample and PVT report available from when the well was drilled, and to pass that composition through the surface process of the field. Another approach could be to use reservoir simulation to estimate current well flowing composition. Another approach could be to use samples of commingled well streams at processing facilities. A clarification about this has been included in the article.

6.There is not enough data validation. All the comparison are based on the proposed SPE case Nr.5 in 1987, which is out of date. Besides, author didn't explain the case in detail, readers cannot reproduce the simulation or calculation.

**Answer by Milan:** I agree, it would be interesting to test the method on other cases to see if it works well. However, I consider showing it working on one case it is sufficient to merit publication.

Regarding the sentence: "SPE case Nr. 5 in 1987, which is out of date", I respectfully disagree with the reviewer. SPE cases are still in wide use today, they do not expire. I don't see this as a serious argument to reject publication.

Some details of the SPE nr. 5 case are given in the appendix. All details are given in the reference: Killough & Kossack, 1987. It is unnecessary to repeat the details here.

All simulation and calculation files are now attached to the delivery for anyone to do verification.

7.Some sentences are very weird, for example Eg. Page 2 line 26: "Results show that the proposed model predicts with reasonable accuracy the well gas-oil ratio…"

**Answer by Milan:** Thank you for the comment. I am sorry to hear you find the sentence weird. We will improve it when preparing for future submissions.

Eg. Page 2 line29: "When using … later time, …" the time here is confusing. It is hard for readers to follow your idea. Please be specific.

**Answer by Milan:** Thank you for the comment. I am sorry to hear you find the sentence confusing. We will improve it when preparing for future submissions.

Eg. Page 9 section 3: "a surface gas and surface oil compositions and respectively…" unreadable.

**Answer by Milan:** Thank you for the comment. I am sorry to hear you find the sentence unreadable. We will improve it when preparing for future submissions.